# A Novel Open Phase Detection Method with Synchronized Phasors for Distribution Systems

Jinlei Xing, Longhua Mu

***Abstract*— An effective open phase detection is one of the unsolved challenges for utilities. Open phase detection is expected to detect the broken power line before it falls to the ground, to prevent the ignition of wildfires. The existing open phase detection methods are normally based on abnormal currents or voltages due to three phase unbalances after open phase events. For a complicated distribution feeder with multiple branches, current-based methods usually have a limited protection range, and voltage-based methods possibly failed due to the recreated voltage on the open phase due to back-feed. This paper presents a new open phase detection method based on the connection impedance of two measuring points, which is calculated with the synchronized phasors. Under an internal open phase condition, the calculated positive-sequence impedance is much larger than the impedance under other conditions, and its sign is negative. This new method is applicable for a distribution feeder with a closed-loop configuration, or DER units connected at the load side. It is almost not impacted by back-feed. This new method only requires point-to-point communication. This method can be easily implemented in the existing recloser controllers or RTUs. It is scalable, cost effective, and reliable.**

## I. INTRODUCTION

OPEN phase condition means the loss of one or two phases due to a broken conductor, loss connection, blown fuse, incorrect switching operations, etc. Open phase conditions are a long-standing concern in distribution systems. An effective open phase detection is one of the unsolved challenges for utilities.

One critical application of open phase detection is the identification of broken conductors. If a broken conductor falls to the ground, it may result in a high-impedance earth fault and can rapidly ignite wildfires due to arc energy or sparking [1], [2]. Conventional earth-fault protection schemes based on current measurements are generally not sensitive enough to detect such high-impedance faults [3], [4]. When a downed conductor is located on the load side, the fault current, resulting from back-feed through distribution transformers, is often too small to be detected. Even if conventional protection eventually operates, the duration of arcing may already be sufficient to initiate wildfire.

Some recently proposed methods based on arcing-current waveform features have not demonstrated sufficient reliability for practical applications. Statistics indicate that broken conductors represent the largest contributor to wildfire incidents associated with electric power facilities [5]. Terms such as “downed conductors” or “falling conductors” refer to the consequences of such failures. For example, the devastating Maui wildfires of 2023 in Hawaii were attributed to broken power lines [6]. Consequently, developing reliable methods to detect broken conductors is a feasible and necessary approach for wildfire mitigation [7]. Utilities aim to detect broken conductors and de-energize affected lines before they contact the ground.

In addition to wildfire risks, downed conductors degrade service reliability and pose serious safety hazards due to potential human contact [8]. Conductors downed on the source side may remain energized due to undetected high-impedance faults, while those on the load side may remain energized due to back-feed through transformers or distributed energy resources (DERs). Furthermore, unloaded $Y_g/\Delta$-connected transformers or three-legged core transformers can restore near-normal voltage on the open phase of the high-voltage side due to electromagnetic coupling [9]. Currently, no effective solution is available that can reliably detect downed conductors under back-feed conditions.

Conventional methods for open phase detection are typically based on phase unbalance, which arises when one or two phases are open. Current-based methods usually rely on negative-sequence currents [10] or residual currents [11], assuming that an open phase results in a significant reduction in load current and corresponding phase unbalance. However, in practical distribution systems where multiple branch lines are connected to a main feeder, the current change depends on the open phase location and load distribution. As a result, current-based methods often provide only partial protection coverage along long feeders [12].

Voltage-based methods detect abnormalities such as undervoltage, rate-of-change of voltage, negative-sequence voltage, or zero-sequence voltage at the load side [13], [14], [15], assuming that an open phase produces a noticeable voltage deviation. In practice, voltage restoration due to back-feed can mask these changes, leading to detection failure. The performance of voltage-based methods is further influenced by transformer configurations, motor loads, and inverter-based DERs. Therefore, such methods are generally considered

Jinlei Xing is with Schneider Electric, Franklin TN37067, USA (e-mail: jinlei.xing@se.com). Longhua Mu is with the College of Electronics & Information Engineering, Tongji University, Shanghai 201804, China (e-mail: lhmu@tongji.edu.cn).

unreliable for open phase detection.

To improve detection performance, some approaches combine multiple measurement quantities, including voltage, current, and impedance [16], [17]. However, these schemes are often complex and rely on measurements from multiple feeder remote terminal units (RTUs). Other techniques based on power quality metrics, voltage distortion [18], or third-harmonic power [19] may be adversely affected by nonlinear loads. Carrier communication–based methods [20] and distributed measurement approaches require multiple monitoring points along the feeder. In addition, most existing methods require a time delay (e.g., 2 s) longer than the clearing time of conventional overcurrent protection to ensure reliability. For wildfire mitigation, much faster operation is required; ideally, the line should be de-energized within 1 s before the broken conductor reaches the ground.

The application of synchronized phasors has improved detection speed in distribution systems [21]. However, most existing methods still rely on abnormal voltage behavior and may fail under back-feed or DER conditions. An impedance-based method using synchronized local and remote measurements has been proposed in [22], where detection is based on the change ratio of calculated impedance. Although effective in principle, this method requires careful threshold tuning, additional blocking logic, and coordination among multiple measurement units. Its operation may also be restricted under light-load conditions, and it may not achieve sufficiently fast tripping for wildfire prevention.

Furthermore, for complex distribution systems with multiple branches, existing impedance-based methods do not provide a practical means of calculating the impedance between two terminals of a feeder section. Installing sufficient measurement points along long feeders is often impractical. While impedance-based protection is relatively straightforward in transmission systems with well-defined terminals [23], its application in distribution systems remains challenging.

This paper proposes a novel method for fast and reliable open-phase detection based on the impedance calculated between two measurement points, referred to as connection impedance. Under normal conditions, the calculated impedance closely matches the actual line impedance. When an open-phase condition occurs, the connection impedance increases significantly and is primarily determined by the equivalent load impedance. Even small current changes can trigger impedance calculation, making the method effective under lightly loaded or unloaded conditions. The main contributions of this paper are as follows:

1) The proposed method is independent of feeder topology and is unaffected by lateral branches.

2) The proposed method requires only point-to-point communication between two terminals.

3) An open phase detection method based on the change of calculated connection impedance is presented. A small current change can trigger impedance calculation, making the method effective under back-feed conditions. This method remains

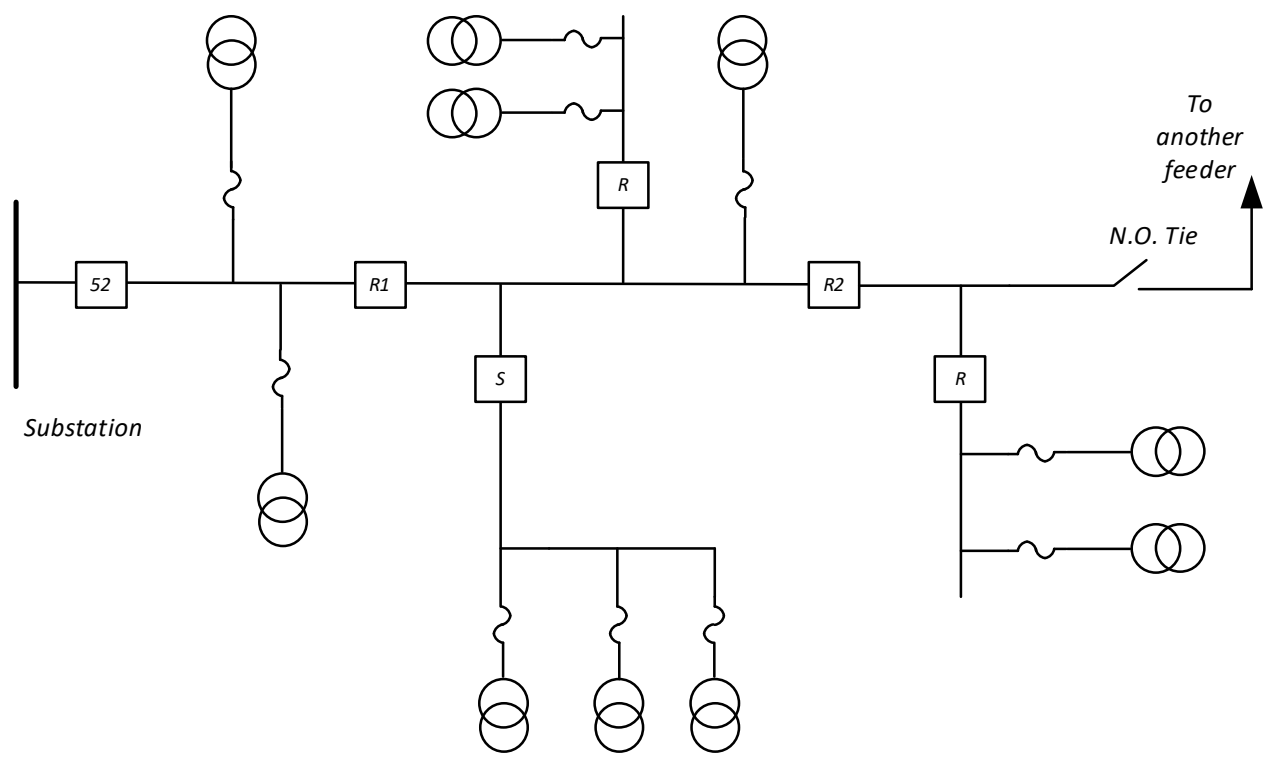

**Fig. 1.** A distribution feeder with multiple lateral taps.

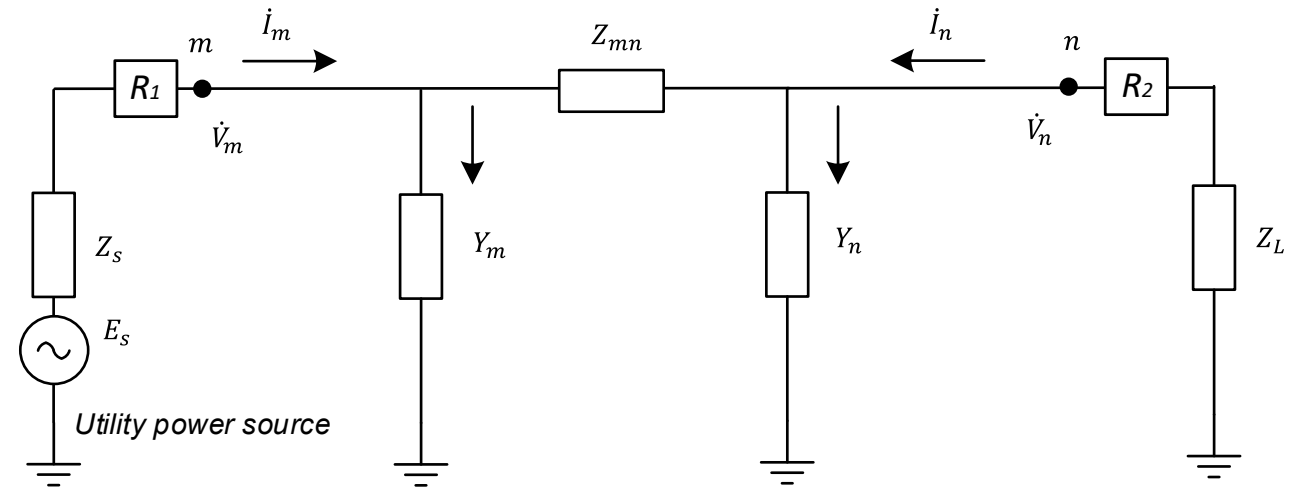

**Fig. 2.** Equivalent circuit of a distribution feeder with multiple lateral taps.

effective even when voltage restoration masks voltage deviations.

This paper is organized as follows. Section II introduces the calculation of connection impedance. Section III presents the theoretical analysis under open phase conditions. Section IV proposes the open phase detection method based on connection impedance. Section V provides simulation results. Section VI discusses practical applications, and Section VII concludes the paper.

## II. Calculation of Connection Impedance

### *A. Simplification of Distribution Feeder*

Distribution feeders deliver power from substations to end users and typically range from several to over one hundred miles in length. In addition, numerous branches—commonly referred to as laterals or lateral taps—extend from the main feeder and form subordinate sub-circuits. As a result, a single distribution feeder often evolves into a complex network, as illustrated in Fig. 1. To ensure effective fault detection and isolation, various protective devices are deployed, including circuit breakers, reclosers, sectionalizing switches, and fuses.

According to [24], the impedance of a main feeder is significantly smaller than the equivalent load impedance in distribution systems. Therefore, a feeder section with multiple lateral taps can be approximated by an equivalent $\Pi$ model, as illustrated in Fig. 2. The tapped loads are represented by two equivalent admittances $Y_m$ and $Y_n$, connected at two terminals of the feeder section. The equivalent line impedance $Z_{mn}$ approximates the actual line impedance [24]. The downstream loads are collectively represented by an equivalent load impedance $Z_L$.

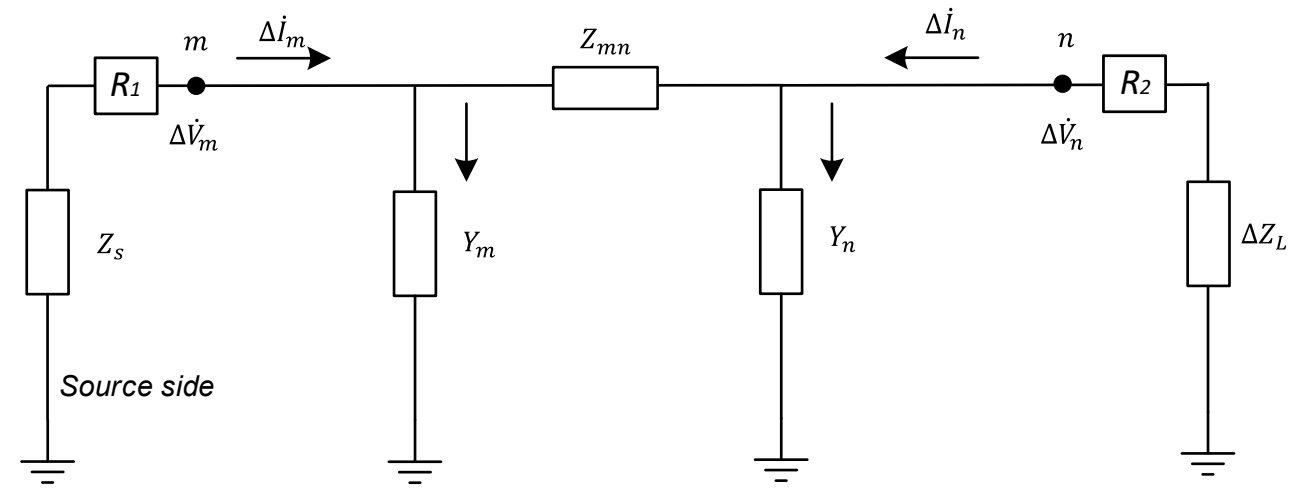

**Fig. 3.** Superimposed equivalent circuit of Fig.2 due to load changes

*B. Connection Impedance*

When the downstream load $Z_L$ varies, the current flowing through the feeder section changes accordingly. In that case, the equivalent line impedance $Z_{mn}$ and the tapped load admittances $Y_m$ and $Y_n$ remain unchanged. Based on this condition, the corresponding superimposed equivalent circuit can be derived, as illustrated in Fig. 3. The resulting changes in voltage and current phasors are evaluated over a time interval $\Delta T$, as defined in (1) and (2).

$$\begin{cases} \Delta\dot{V}_m = \dot{V}_m(t) - \dot{V}_m(t-\Delta T) \\ \Delta\dot{V}_n = \dot{V}_n(t) - \dot{V}_n(t-\Delta T) \end{cases} \tag{1}$$

$$\begin{cases} \Delta\dot{I}_m = \dot{I}_m(t) - \dot{I}_m(t-\Delta T) \\ \Delta\dot{I}_n = \dot{I}_n(t) - \dot{I}_n(t-\Delta T) \end{cases} \tag{2}$$

Equation (3) is derived based on the equivalent circuits shown in Fig. 2 and Fig. 3. Using this relationship, the equivalent line impedance can be determined. The calculated impedance $Z_{mn}$ in (4) approximates the actual line impedance under normal conditions.

$$\begin{cases} \dot{V}_m - \dot{V}_n = (\dot{I}_m - Y_m\dot{V}_m)Z_{mn} \\ \Delta\dot{V}_m - \Delta\dot{V}_n = (\Delta\dot{I}_m - Y_m\Delta\dot{V}_m)Z_{mn} \end{cases} \tag{3}$$

$$Z_{mn} = \frac{\Delta\dot{V}_m\dot{V}_n - \Delta\dot{V}_n\dot{V}_m}{\Delta\dot{I}_m\dot{V}_m - \Delta\dot{V}_m\dot{I}_m} \tag{4}$$

The impedance $Z_{mn}$ is defined as the connection impedance between two measurement points $m$ and $n$, regardless of the number of lateral branches between them. The parameter $\Delta T$ represents the calculation time window and determines the number of superimposed samples available for impedance estimation following a current change.

According to (4), connection impedances can be calculated from both the local and remote terminals when a current change is detected, as given by:

$$Z_{LR} = \frac{\Delta\dot{V}_L\dot{V}_R - \Delta\dot{V}_R\dot{V}_L}{\Delta\dot{I}_L\dot{V}_L - \Delta\dot{V}_L\dot{I}_L} \tag{5}$$

where:

$\dot{V}_L$ and $\dot{I}_L$ are the local voltage and current phasors;
$\Delta\dot{V}_L$ and $\Delta\dot{I}_L$ are variations over the interval $\Delta T$;
$\dot{V}_R$ and $\dot{I}_R$ are the remote voltage and current phasors;
$\Delta\dot{V}_R$ and $\Delta\dot{I}_R$ are variations over the interval $\Delta T$.

The voltage and current phasors may be expressed either as phase quantities or as symmetrical components.

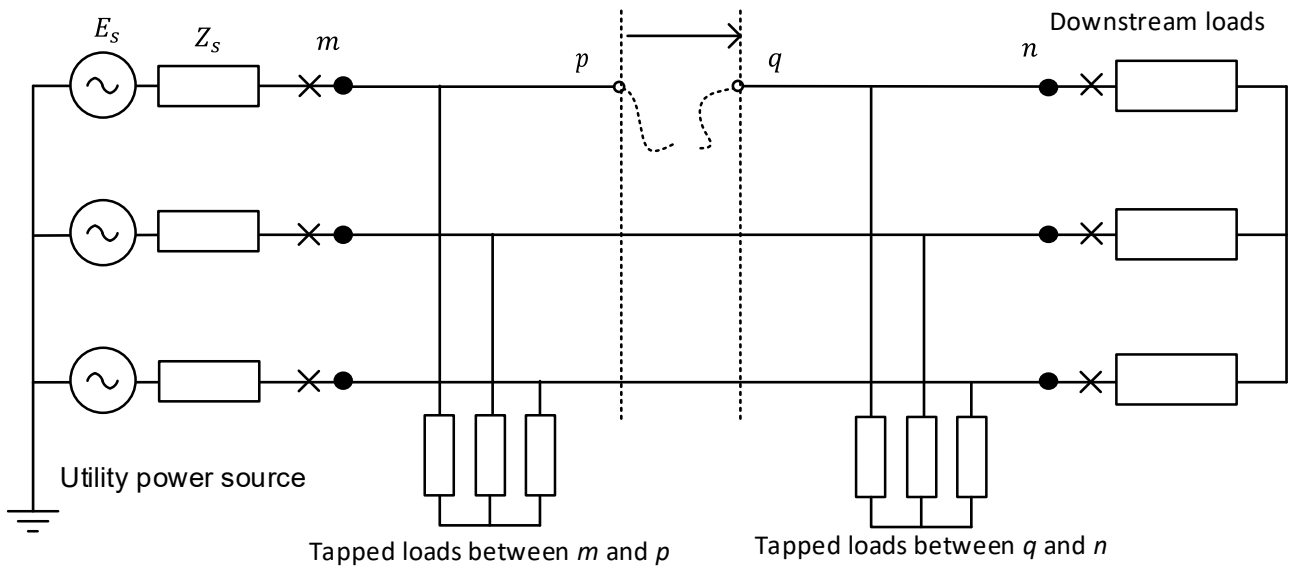

**Fig. 4.** Open phase analysis (take phase A open for example)

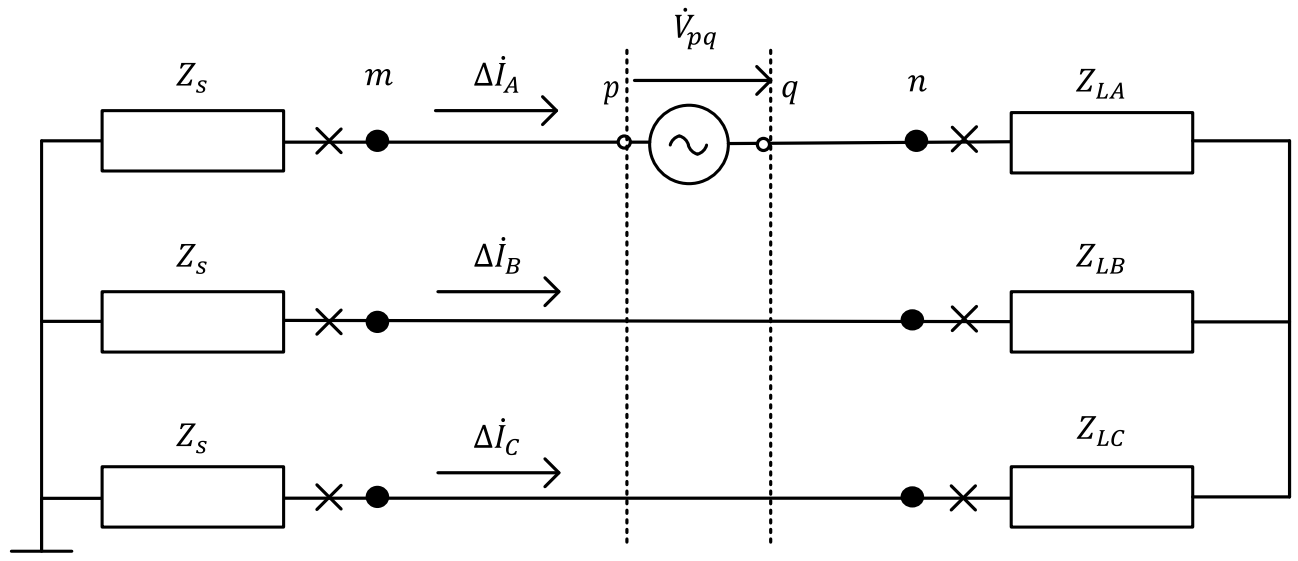

**Fig. 5.** Superimposed equivalent circuit of phase A open

*C. Connection Impedance Application*

For practical implementation of the connection impedance concept, it is necessary to analyze the connection impedance under different system conditions.

Let $Z_{ext}$, $Z_{int}$, and $Z_{op}$ denote the connection impedance under an external event, an internal event, and an internal open phase condition, respectively. For external disturbances or system faults located outside the monitored section, the calculated connection impedance $Z_{ext}$ approximates the actual line impedance [24]. In contrast, for internal events or faults within the monitored section, the calculated connection impedance $Z_{int}$ is typically smaller than the line impedance, as discussed in [24]. The following section will analyze the connection impedance under open phase conditions.

## III. Open Phase Analysis

Consider a three-wire distribution system in which phase A is open, as illustrated in Fig. 4. Tapped load branches are present both upstream and downstream of the open phase location. Measurement point $m$ is connected to a strong utility source, whereas measurement point $n$ is located on the load side and may be connected to a weak source, such as an inverter-based DER.

*A. Calculation from Source Side*

At measurement point $m$, a current variation is detected at the instant of open phase. The voltage variation at the power-source side is negligible and can be approximated as (6).

$$\Delta\dot{V}_{(source\ side)} = \Delta\dot{V}_{m.A} \approx 0 \tag{6}$$

Then, the connection impedance between points $m$ and $n$ for a phase-A open condition is expressed as in (7).

$$Z_{mn.op.A} \approx \frac{-\Delta\dot{V}_{n.A}\dot{V}_{m.A}}{\Delta\dot{I}_{m.A}\dot{V}_{m.A}} \approx \frac{-\Delta\dot{V}_{n.A}}{\Delta\dot{I}_{m.A}} \tag{7}$$

An equivalent voltage source can be introduced to represent the open phase condition, as illustrated in Fig. 5. The voltage source $V_{pq}$ corresponds to the voltage difference between the broken terminal $p$ at the source side and the broken terminal $q$ at the load side. The tapped loads between points $m$ and $p$ can be neglected, as their equivalent impedance is significantly larger than the equivalent impedance $Z_S$ of the power source. The tapped loads between points $q$ and $n$ , and the downstream loads are indicated by equivalent phase impedance $Z_{LA}$, $Z_{LB}$, $Z_{LC}$ as the downstream loads of an open phase location.

Assume that three-phase loads are approximately balanced. Based on the equivalent circuit shown in Fig. 5, the total impedance $Z_{L,op,A}$ associated with the equivalent voltage source $V_{pq}$ under a single-phase open condition can be expressed as (8).

$$Z_{L.op.A} = (Z_s + Z_{LA}) + (Z_s + Z_{LB})||(Z_s + Z_{LC}) \approx 1.5(Z_s + Z_{LA}) \quad (8)$$

The connection impedance $Z_{L.op.A}$ can also be directly estimated from the superimposed quantities as (9).

$$Z_{L.op.A} \approx \frac{\dot{V}_{pq}}{\Delta \dot{I}_{m.A}} \quad (9)$$

The voltage variation at the power-source side is negligible. Then the equivalent voltage source can be approximated as the negative of the load side voltage variation, as (10).

$$\dot{V}_{pq}(t) \approx -\Delta \dot{V}_{n.A}(t) \quad (10)$$

The following conclusion can be drawn. When an open phase condition occurs between the two measurement points, the connection impedance calculated from the source side is approximated as the total open phase equivalent impedance, which is primarily determined by the downstream load impedance. Moreover, the calculated connection impedance has a negative sign, as (11).

$$Z_{mn.op.A} \approx -Z_{L.op.A} \approx -1.5(Z_s + Z_{LA}) \quad (11)$$

The above analysis is based on a three-wire, ungrounded distribution system. For a four-wire, multi-grounded system, the connection impedance under a single-phase open condition can be expressed as (12).

$$Z_{mn.op.A} \approx -Z_{L.op.A} \approx -(Z_s + Z_{LA}) \quad (12)$$

Therefore, under an internal single-phase open condition, the calculated connection impedance is dominated by the equivalent load impedance and is significantly larger than the line impedance in distribution systems.

*B. Back-Feed via $Yg/\Delta$ Transformer*

A $Y_g/\Delta$-connected transformer under no-load condition, in theory, can restore nearly normal voltage on the high-voltage side due to electromagnetic coupling [9]. This scenario represents a worst-case condition, as open phase detection methods based solely on voltage abnormalities may fail to identify the fault.

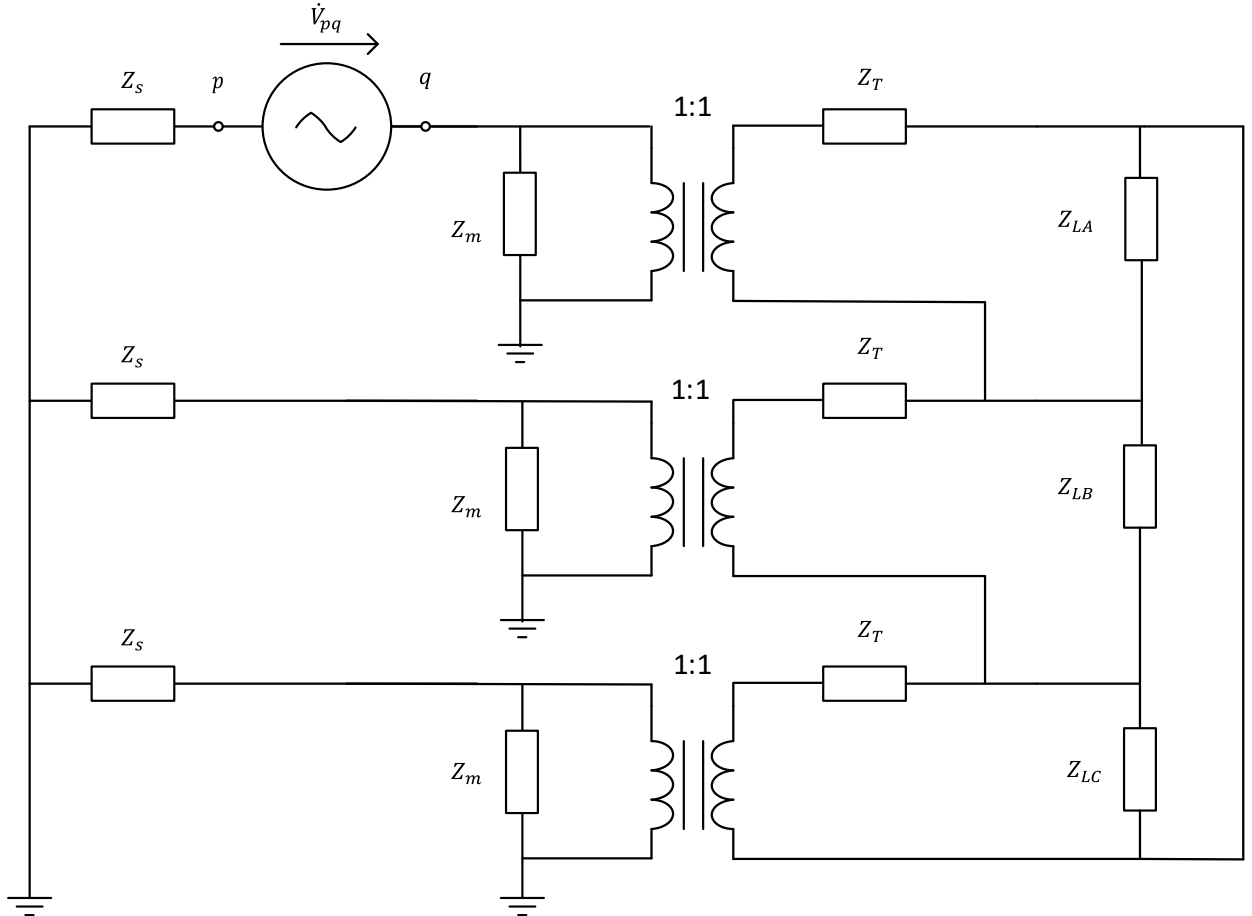

**Fig. 6.** Superimposed equivalent circuit with $Yg/\Delta$-connected transformer

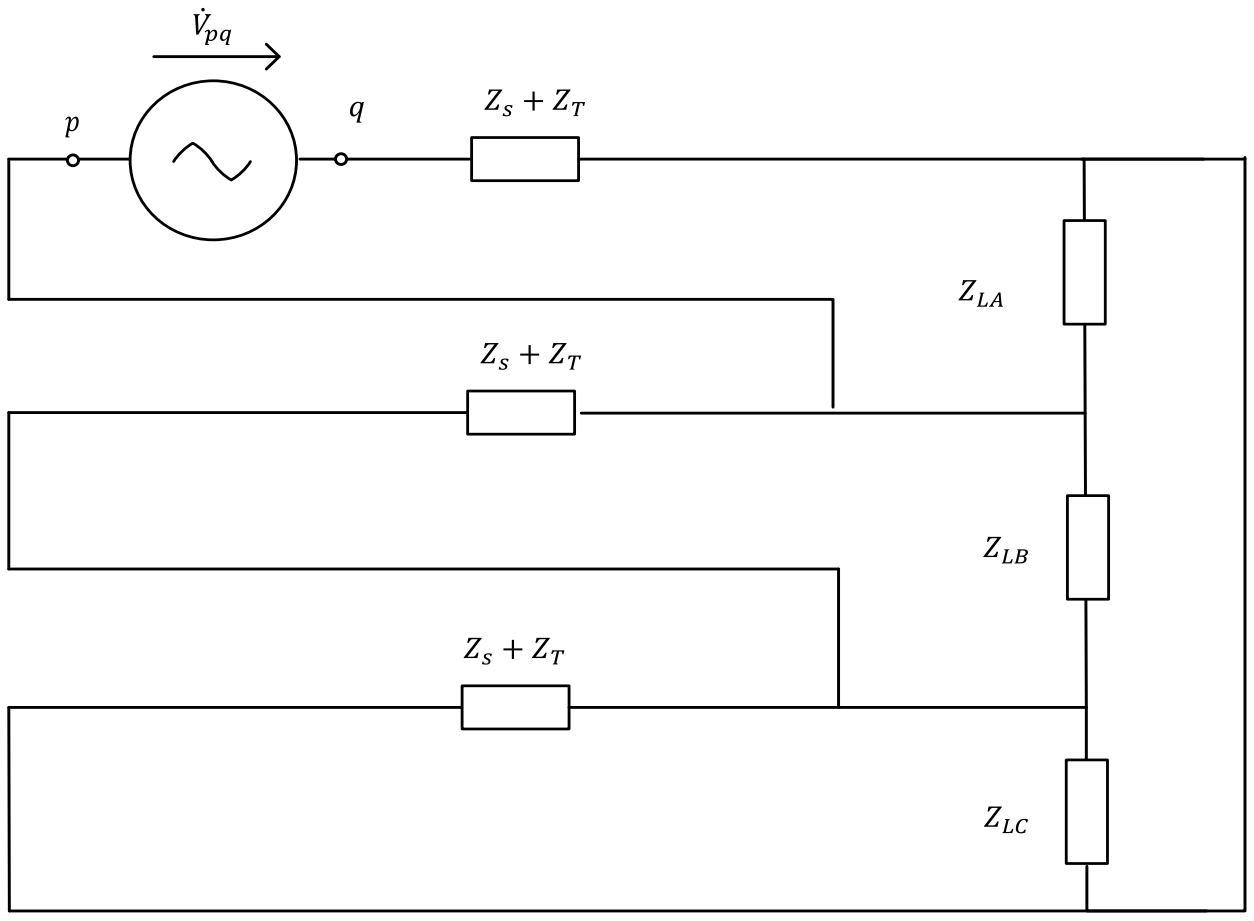

**Fig. 7.** Simplification of Fig.6

Similar to Fig. 5, the superimposed equivalent circuit for a phase-A open condition with a single $Y_g/\Delta$ -connected transformer is shown in Fig. 6. To simplify the analysis, all transformer loads are referred to the HV side, and the transformer turns ratio is assumed to be unity.

In Fig. 6, $Z_s$ denotes the equivalent impedance of the utility source, $Z_m$ represents the transformer magnetizing impedance, and $Z_T$ is the transformer leakage impedance. The load impedances $Z_{LA}$ , $Z_{LB}$ , and $Z_{LC}$ correspond to the equivalent phase loads. The resulting equivalent circuit used for connection impedance analysis under an open phase condition is shown in Fig. 7, which highlights the impact of the low-voltage (LV) Δ-connected winding, and ignores the transformer excitation impedance.

For distribution transformers, the transformer impedance typically ranges from 2% to 6% of the rated load impedance. For a light-loaded transformer, its equivalent load impedance is significantly larger than both the transformer impedance and the equivalent source impedance. Consequently, the equivalent load impedance $Z_{LA}$, $Z_{LB}$, $Z_{LC}$ can be neglected in the analysis with Fig. 7. Under these assumptions, the connection impedance for phase-A open with a single $Y_g/\Delta$-connected transformer can be expressed as (13).

$$Z_{L.op.A} \approx 3(Z_s + Z_T) \tag{13}$$

If multiple $Y_g/\Delta$-connected distribution transformers are installed downstream of the open point, they effectively operate in parallel. As a result of parallel impedance, the calculated connection impedance based on phase quantities decreases accordingly.

### *C. Calculation Based on Positive-Sequence Quantities*

The above analysis is based on phase quantities, namely current phasors and phase-voltage phasors, for connection impedance calculation. The impedance values obtained under open phase conditions in (11) and (12) differ between three-wire ungrounded systems and four-wire multi-grounded systems. In addition, multiple downstream $Y_g/\Delta$-connected transformers reduce the calculated connection impedance. Furthermore, system configurations, such as load types (three-phase and single-phase), grounding configurations, and DERs interconnections, affect the phase quantities and consequently, the connection impedance calculation results.

To mitigate the influence of system configurations, it is recommended to use positive-sequence quantities for the connection impedance calculation in (5). Both three-wire ungrounded systems and four-wire multi-grounded systems share the same symmetrical component network. As illustrated in Fig. 8, $Z_{S1}$, $Z_{S2}$, $Z_{S0}$ are the positive, negative and zero sequence impedances of the power source; $Z_{L1}$, $Z_{L2}$, $Z_{L0}$ are the equivalent positive, negative and zero sequence impedances downstream loads of the open phase location.

The positive-sequence voltage variation at the power source side is negligible and can be approximated as (14).

$$\Delta\dot{V}_{(source\ side)} = \Delta\dot{V}_{m.1} \approx 0 \tag{14}$$

According to (5) and (14), the positive-sequence connection impedance between points $m$ and $n$ for a phase-A open condition is given by (15).

$$Z_{mn.op.1} \approx \frac{-\Delta\dot{V}_{n.1}\dot{V}_{m.1}}{\Delta\dot{I}_{m.1}\dot{V}_{m.1}} \approx \frac{-\Delta\dot{V}_{n.1}}{\Delta\dot{I}_{m.1}} \tag{15}$$

The relationship between the equivalent voltage and the load side voltage variation can be obtained as (16).

$$\dot{V}_{pq.1}(t) \approx -\Delta\dot{V}_{n.1}(t) \tag{16}$$

The positive-sequence connection impedance can be expressed as (17).

$$Z_{mn.op.1} \approx \frac{\Delta\dot{V}_{pq.1}}{\Delta\dot{I}_{m.1}} = -Z_{L.op.1} \approx -(Z_{s1} + Z_{L1}) \tag{17}$$

The calculated positive-sequence connection impedance, under an internal single-phase open condition, is expressed in (17). It is primarily determined by the downstream load positive-sequence impedance, which normally is much larger than the line positive-sequence impedance. Compared with calculations based on phase quantities in (11), (12) and (13), the positive-sequence connection impedance under an open phase condition, has a unified expression in (17), no matter system configurations. It will effectively simplify the engineering applications of the connection impedance for open phase detection, without the consideration for system configurations.

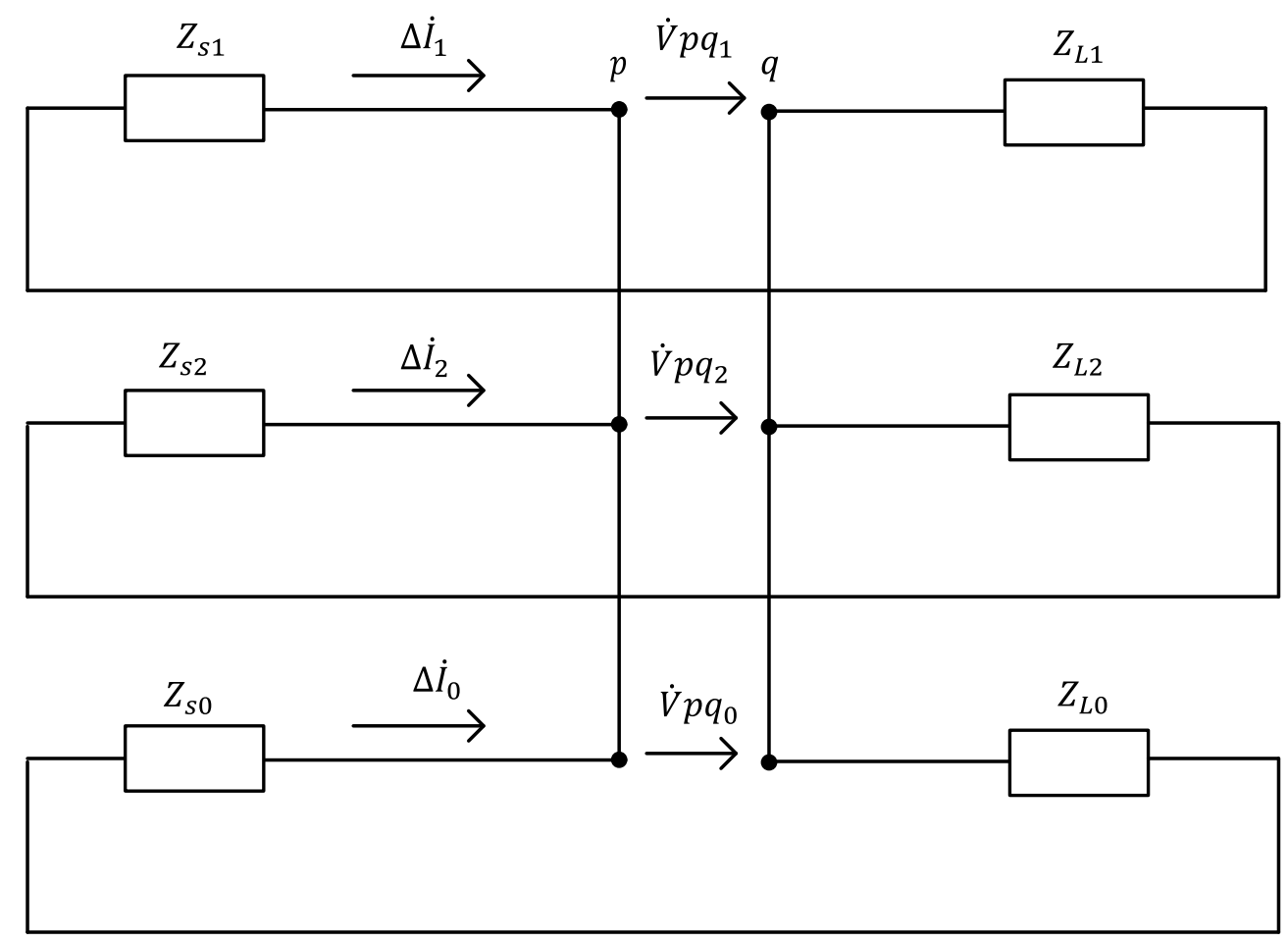

**Fig. 8.** Symmetrical component networks for superimposed equivalent circuit of phase A open

### *D. Calculation from Load Side*

At measurement point $n$, a current variation is also detected at the instant of the conductor break. Based on positive-sequence quantities, the connection impedance between points $n$ and $m$ under a phase-A open condition can be expressed as:

$$Z_{nm.op.1} \approx \frac{\Delta\dot{V}_{n.1}\dot{V}_{m.1}}{\Delta\dot{I}_{n.1}\dot{V}_{n.1} - \Delta\dot{V}_{n.1}\dot{I}_{n.1}} \tag{18}$$

The variations in positive-sequence current and voltage can be represented as proportional to their steady state values:

$$\begin{cases} \Delta\dot{I}_{n.1} = \alpha\dot{I}_{n.1} \\ \Delta\dot{V}_{n.1} = \beta\dot{V}_{n.1} \end{cases} \tag{19}$$

Substituting (19) into (18):

$$Z_{nm.op.1} \approx \frac{\beta\dot{V}_{n.1}\dot{V}_{m.1}}{\alpha\dot{I}_{n.1}\dot{V}_{n.1} - \beta\dot{V}_{n.1}\dot{I}_{n.1}} = \frac{\beta}{\alpha - \beta}\left(\frac{\dot{V}_{m.1}}{\dot{I}_{n.1}}\right) \tag{20}$$

From the symmetrical-component network analysis, it follows that:

$$\frac{\dot{V}_{n.1}}{\dot{I}_{n.1}} \approx \frac{\Delta\dot{V}_{n.1}}{\Delta\dot{I}_{n.1}} \approx -(Z_{s1} + Z_{L1}) \tag{21}$$

which implies that $\alpha \approx \beta$.

Consequently, the denominator in (20) approaches zero, resulting in a very large, calculated value of positive-sequence impedance when evaluated from the load side under an internal open phase condition. In practical applications, measurement noise and numerical errors further affect the stability of this calculation. As a result, the impedance estimated from the load side is typically unreliable and exhibits significant fluctuations.

In contrast, the positive-sequence connection impedance calculated from source side has a clear physical interpretation,

corresponding to the equivalent positive-sequence impedance of the system, and consistently exhibits a negative sign under open phase conditions. Therefore, the source side impedance is preferred for open phase detection in the subsequent analysis.

### *E. Distribution Feeder Connected to Multiple Power Sources*

With the increasing penetration of DERs, a larger number of DER units are connected to distribution feeders at the load side. Most of the DER units are inverter-based and operate in a grid-following mode. Such inverters typically behave as controlled current sources and exhibit relatively higher positive-sequence impedance. Under an internal open phase condition, the calculated positive-sequence connection impedance remains significantly larger than the line impedance, even in the presence of multiple DERs.

Radial and open-loop distribution systems are widely used; however, closed-loop configurations are also employed in some utilities. A typical closed-loop configuration is illustrated in Fig. 9, where two feeders originating from the same substation bus are interconnected through a normally open tie breaker. When the tie breaker is closed for a closed-loop operation, each feeder can be energized from both directions. In that case, the equivalent load impedance is connected in parallel with the source impedance, and its contribution can be neglected. As a result, the calculated positive-sequence connection impedance under an internal open phase condition is approximated to the sum of the source impedance and the line impedance. Therefore, a lower impedance threshold should be adopted for open phase detection in systems with closed-loop operation.

## IV. Open phase Detection Method

Based on the analysis presented in Section III, the connection impedance can be calculated from both the power-source side and the load side using either phase quantities or positive-sequence quantities. Among these, the source side calculation is preferred.

The phase-segregated connection impedance under an open phase condition is primarily determined by the downstream equivalent load impedance; however, downstream $Y_g/\Delta$-connected transformers reduce the calculated value, as shown in (13). In contrast, the positive-sequence connection impedance under an open phase condition is independent of distribution system configurations and is approximated to the downstream load positive-sequence impedance. Nevertheless, $Y_g/\Delta$-connected transformers can restore the phase voltage under open phase conditions, resulting in a reduced variation in positive-sequence current that depends on downstream load conditions. Therefore, for open phase detection, the positive-sequence impedance is preferred when the variation in positive-sequence current exceeds a predefined threshold.

The impedance calculation is initiated upon detection of positive-sequence current variation. The first detection criterion ensures sufficient measurement accuracy as (22). The current threshold $I_{\text{set}}$ should be slightly higher than the current measurement error.

$$\left|\Delta \dot{I}_{m.1}\right| \geq I_{set} \tag{22}$$

To improve reliability, the impedance calculation should be performed at the source side, where the positive-sequence voltage variation is negligible, as described in (14). This leads to the second criterion in (23). The voltage threshold $V_{\text{set}}$ should exceed the measurement error. Therefore, a fixed setting of 5% the rated voltage is suggested.

$$\left|\Delta \dot{V}_{m.1}\right| \leq V_{set} \tag{23}$$

In systems with multiple sources or in cases where voltage is restored due to back-feed through downstream transformers, both terminals may satisfy this criterion. In such scenarios, connection impedance calculation and open phase detection can be performed independently at each terminal.

The third criterion is based on the magnitude of the calculated connection impedance. Since the positive-sequence connection impedance under an open phase condition is dominated by the equivalent load impedance and is significantly larger than the line impedance, the following condition is applied:

$$|Z_{mn.op.1}| \geq Z_{set} \tag{24}$$

The impedance threshold $Z_{set}$ is used to distinguish internal open phase conditions from other system events, such as external faults or load variations or internal faults. This threshold should be selected to be significantly larger than the line impedance $Z_{line}$ between the two measurement points as (25). $Z_{mn.ext.1}$ and $Z_{mn.int.1}$ are the calculated positive-sequence connection impedances under external events and internal events respectively.

$$\begin{cases} |Z_{mn.ext.1}| \approx Z_{line} \ll Z_{set} \\ |Z_{mn.int.1}| \leq Z_{line} \ll Z_{set} \end{cases} \tag{25}$$

For radial or open-loop distribution feeders, the calculated positive-sequence connection impedance under an open phase condition is significantly larger than the line impedance. By default, the impedance setting can be simply configured as 5Ω, larger than 20km line impedance. For closed-loop distribution systems, it requires exact calculations for the impedance setting. The line section for open phase detection shall be much shorter than the total line length. The impedance setting shall be less than the total line impedance but more than the line section's impedance.

Finally, the negative sign of the calculated impedance serves as an important directional indicator and enhances detection reliability, particularly in closed-loop configurations.

## V. Simulations

### *A. Parameters for Simulations*

As shown in Fig. 9, the simulated distribution feeder in Matlab Simulink is a four-wire, multi-grounded system. The system voltage is 12 kV, and the source impedance is 0.9 Ω. The positive-sequence impedance of the overhead line is 0.13Ω/km. It is divided into multiple sections by Reclosers R1 to R5. The parameters of each line section are as follows:

- Section (R1, R2): 3 km, impedance 0.39Ω
- Section (R2, R3): 5 km, impedance 0.85Ω
- Section (R3, R4): 5 km, impedance 0.72Ω
- Section (R4, R5): 10 km, impedance 1.3Ω

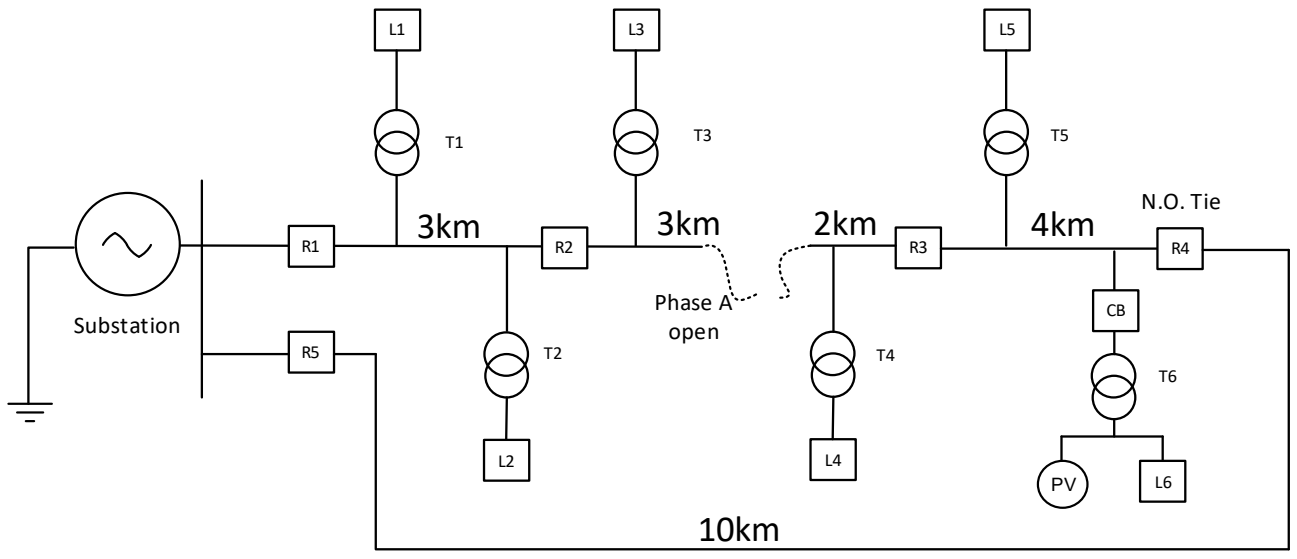

**Fig. 9.** Distribution feeder for open phase simulation

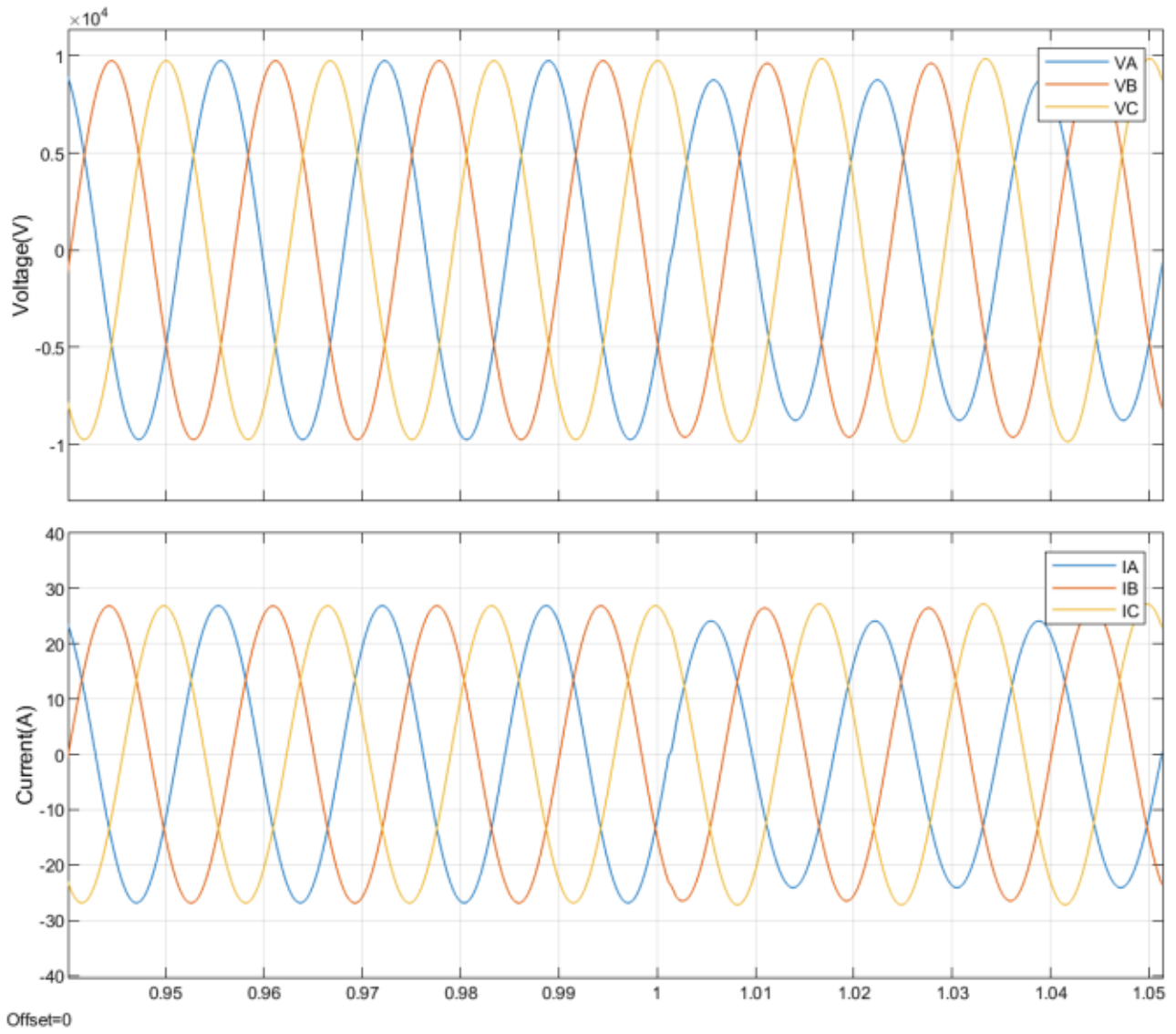

**Fig. 10.** Voltages and currents at R3 before and after Phase A open

A photovoltaic (PV) system is connected on the load side between R3 and R4. All distribution transformers are rated at 1 MVA, 12 kV/0.48 kV, with a short-circuit impedance of 6% (corresponding to 8.6Ω). Transformer configurations and loads are as follows:

- T1: $Y_g/Y_g$, 600 kVA
- T2: $\Delta/Y_g$, 500 kVA
- T3: $Y_g/Y_g$, 800 kVA
- T4: $Y_g/\Delta$, 100 kVA
- T5: $Y_g/Y_g$, 200 kVA
- T6: $Y_g/Y_g$, 200 kVA

The proposed open phase detection method is implemented in each recloser controller. Each feeder section is monitored by a pair of reclosers acting as synchronized measurement points, exchanging phasor data.

An open phase condition is simulated between R2 and R3, where phase A is opened at $t = 1.0$s. The calculation window is set to $\Delta T = 1.0$s, allowing the connection impedance to be evaluated over the interval from 1.0 s to 2.0 s.

The total downstream load is 500 kW, corresponding to an equivalent impedance of approximately 288Ω. The current-change threshold for initiating impedance calculation is set to 1 A, while the voltage-change threshold for identifying the source side is set to 5%.

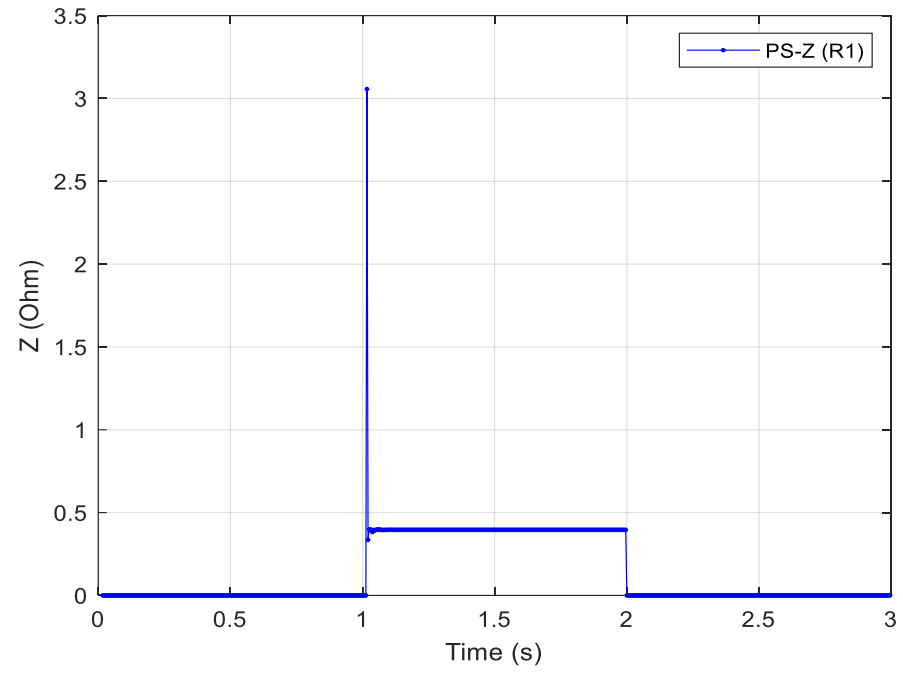

**Fig. 11.** Calculated connection impedance at R1 under an external open phase

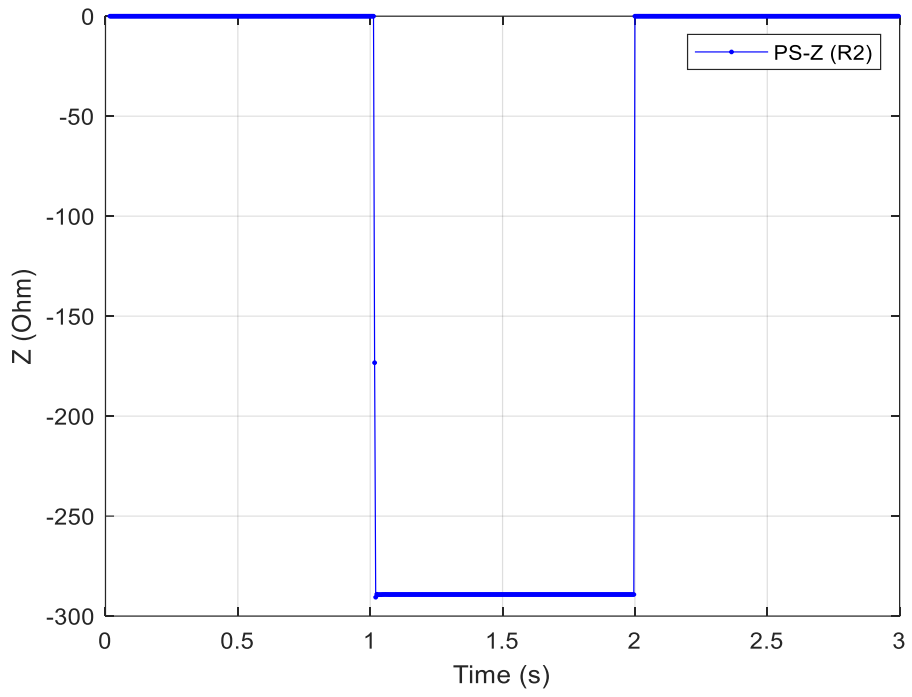

**Fig. 12.** Calculated connection impedance under an internal phase A open at R2

### *B. Single power source with open-loop configuration*

In this simulation scenario, the normally open tie switch R4 remains open, and the load side PV system is disconnected. Under these conditions, transformer T4, configured as $Y_g/\Delta$, restores the phase-A voltage downstream of the open phase location. As shown in Fig. 10, the voltage deviation following phase A open condition is about 10%, and the corresponding current change in phase A is about 3A, which triggers the calculation of positive-sequence connection impedance for open phase detection.

For line section (R1, R2), the simulated scenario corresponds to an external open phase event. In this case, the connection impedance calculations are triggered at both measurement points R1 and R2. According to the analysis in Section III, the calculated connection impedance between these two points is expected to approximate the line impedance. As shown in Fig. 11, the calculated connection impedance at R1 is 0.396Ω, which is nearly identical to the actual line impedance. The same value is obtained at R2. These results are consistent with the theoretical analysis presented in Section III and validate the expected behavior for external events.

For line section (R2, R3), the simulated scenario represents an internal open phase event. The voltage variation at the load side terminal R3 is about 10%, more than 5% as shown in Fig.10. Therefore, the impedance calculation is not triggered at R3. According to the analysis in Section III, the calculated positive-sequence connection impedance at R2 should approximate the downstream equivalent load impedance. In this scenario, the total downstream load is 500 kW, corresponding to an equivalent impedance 288Ω in a 12 kV system. As shown in Fig. 12, the calculated positive-sequence connection impedance is 289.1Ω, which closely matches the expected equivalent load impedance.

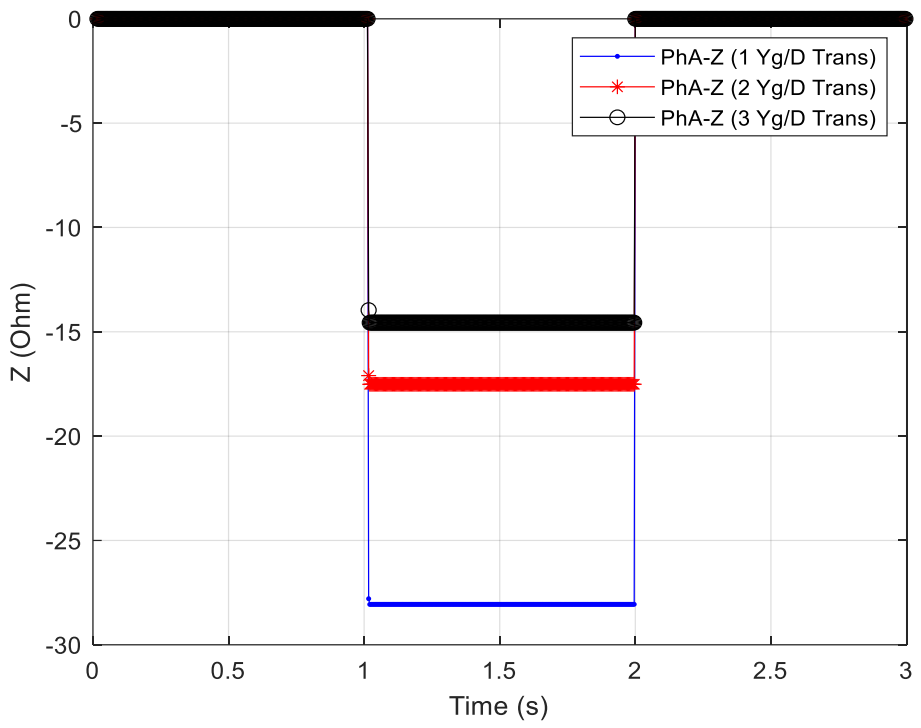

**Fig. 13.** Phase A connection impedance at R2 with (1~3 $Y_g/\Delta$ transformers)

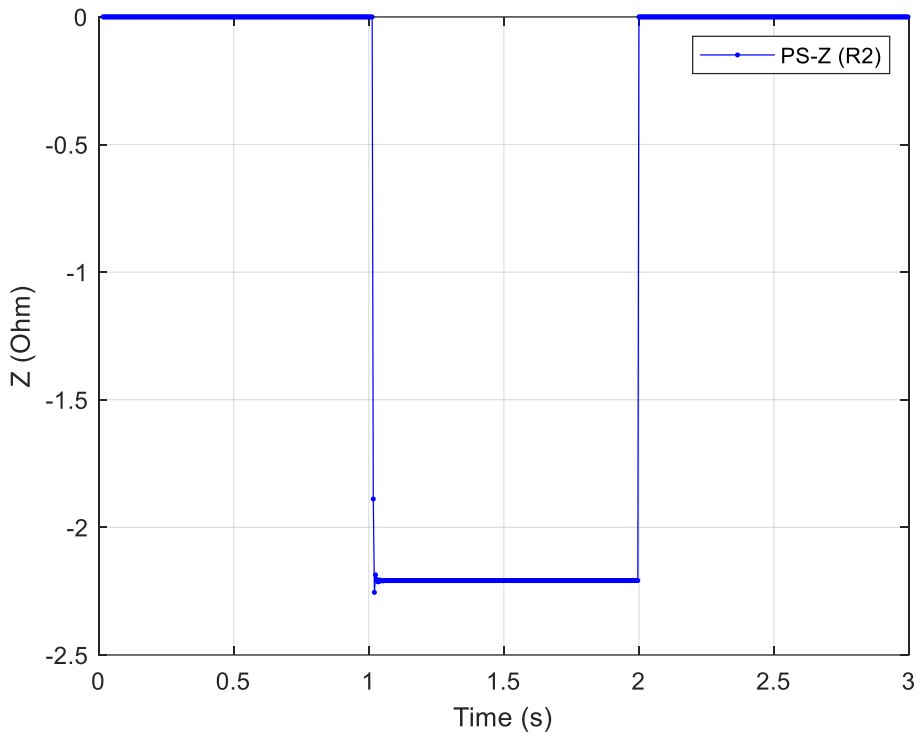

**Fig. 14.** Positive-sequence connection impedance at R2 (closed loop)

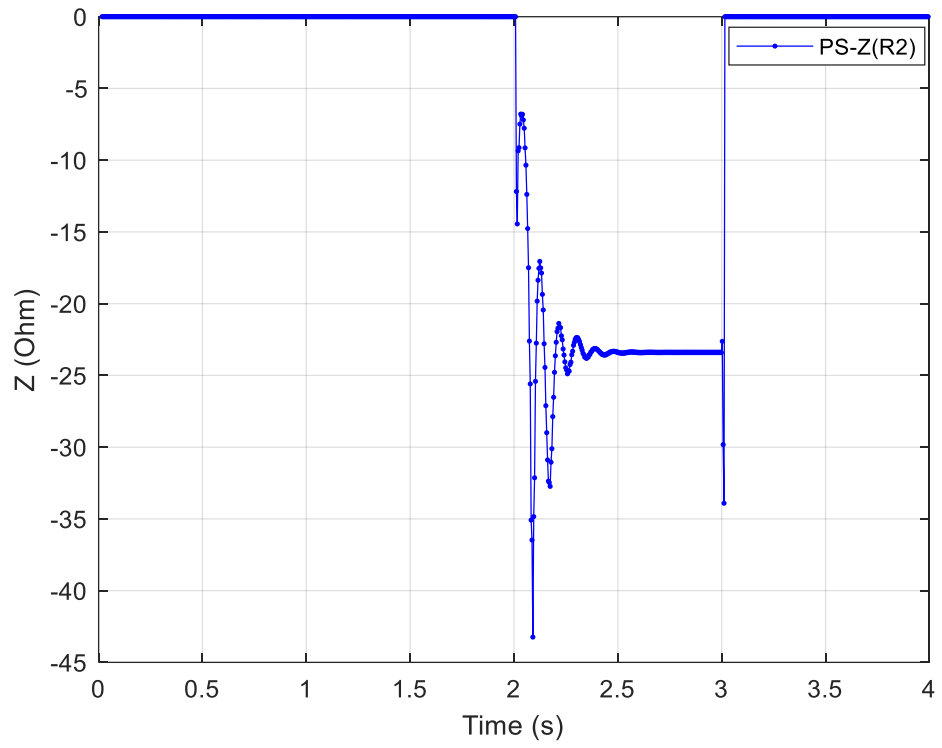

**Fig. 15.** Positive-sequence connection impedance at R2 (PV connected)

When phase quantities are used to calculate phase A connection impedance, the presence of $Y_g/\Delta$-connected transformers affect the resulting connection impedance, as discussed in Section III. The calculated impedance depends on both the transformer impedance and the number of downstream $Y_g/\Delta$-connected transformers.

In this simulation case, the transformer impedance is 8.6Ω, and the source impedance is 0.9Ω. Based on the theoretical analysis in Section III, the connection impedance under an open phase condition is expected to be about 28.5Ω according to (15). As shown in Fig. 13, the calculated phase A connection impedance is 28.1Ω, which is in line with the theoretical value. When multiple $Y_g/\Delta$-connected transformers are present downstream of the open phase location, their parallel effect reduces the equivalent impedance. Specifically, the calculated phase A connection impedance decreases to 18.3Ω with two transformers and to 14.5Ω with three transformers, as shown in Fig. 13.

### C. Two Power Sources with Closed-loop Configuration

In this simulation scenario, the normally open tie switch at R4 is closed, and the load side PV system is disconnected. The system therefore operates in a closed-loop configuration, with two power sources supplied from the same utility substation.

For the open phase event occurring within the line section $(R2, R3)$, both R2 and R3 act as source side terminals, which leads to the calculation of positive-sequence impedance at both locations. According to the analysis in Section III, when a second power source is present at the load side, the equivalent load impedance is effectively connected in parallel with the second source impedance and can therefore be neglected. In this simulation case, since both sources originate from the same substation bus in a closed-loop configuration, the equivalent source impedance can also be considered negligible.

As a result, the calculated positive-sequence connection impedance is primarily determined by the surrounding line sections, including (R1, R2) and (R3, R4). As shown in Fig. 14, the calculated positive-sequence connection impedance at R2 is 2.21Ω, and the same value is obtained at R3. These results indicate that, for closed-loop configurations, the impedance magnitude under an open phase condition is significantly influenced by the line impedance rather than the load impedance. Therefore, proper placement of measurement points and careful selection of impedance thresholds are essential. In particular, the impedance threshold must be set lower than the total line impedance of the monitored section to ensure reliable open phase detection.

### D. DER Connected at Load Side

In this simulation scenario, the normally open tie switch at R4 remains open, while a 700 kW PV system is connected to transformer T6. The open phase condition is located between R2 and R3. Considering the dynamic response of the PV system, the phase A open condition is introduced at $t = 2.0$s, with a calculation window of $\Delta T = 1.0$ s. Consequently, the connection impedance can be evaluated over the interval from 2.0 s to 3.0 s.

With the PV system connected downstream of the open phase location, the $Y_g/\Delta$-connected transformer T4 cannot fully restore the phase A voltage. In this case, the calculated positive-sequence connection impedance at R2 is primarily determined by the positive-sequence impedance of the PV system, which is significantly larger than the line impedance.

As shown in Fig. 15, the steady state value of the calculated positive-sequence connection impedance at R2 is approximately 23.4Ω. The result also captures the transient response of the PV system at the instant of the phase A open event.

### E. Lab Simulation Tests on RTU devices

The proposed open phase detection in this paper has been implemented in a commercialized RTU device, which is widely applied in distributions systems. Two RTU devices are connected and synchronized by 4G network for lab simulation tests. One purpose of the lab simulation tests is to verify the 4G communication latency.

The settings of open phase detection for lab simulation tests are shown in Table I. These settings are all default settings, which are applicable for most of the practical engineering applications. For each test case in Table II, Matlab Simulink is applied to generate the related testing waveforms which are injected into two RTU devices by a relay test set. The test results show that the tripping time for open phase detection is less than 1.0s even if 300ms operation time delay is applied to ensure reliability. The 4G communication latency is not a concern for practical applications because the conservative estimated falling time is 1.06s for a broken conductor reaches the ground [22]. The proposed open phase detection keeps stable for external open phase events and internal earth faults.

TABLE I.
Settings of Open Phase Detection for Lab Simulation Tests

| Setting name | Setting value |
|---|---|
| Current change threshold | 0.01 p.u |
| Calculation time window $\Delta T$ | 800 ms |
| Impedance threshold | 5 Ω |
| Operation time delay | 300 ms |

TABLE II.
Open Phase Detection Test Results of Lab Simulation Tests

| Test cases | Average trip time (repeat 10 tests) |
|---|---|
| Internal open phase | 670 ms |
| External open phase | Not trip |
| Internal earth fault | Not trip |
| Internal open phase with back-feed | 660 ms |
| Internal open phase with PV connected | 690 ms |

## VI. Discussions

The core principle of the proposed open phase detection is the evaluation of the electrical connection impedance between two measurement points. Regardless of system topology, when all three phases are intact, the connection impedance remains small and is approximately equal to the line impedance. Moreover, it is largely unaffected by the presence of multiple lateral branches. In contrast, when any phase is open, the connection impedance increases significantly and becomes much larger than the line impedance between the two measurement points.

The implementation of the proposed open phase detection is straightforward. It requires only point-to-point communication to exchange synchronized phasor measurements between two monitoring locations, which can be achieved using 4G networks or direct fiber-optic links. The method can be readily integrated into existing recloser controllers or RTUs. For long distribution feeders with multiple sections and branches, the proposed method does not require a larger number of measurement points, making it highly scalable. At a minimum, only two measurement points are needed. For example, in the feeder shown in Fig. 9, if selectivity is not required, the method can be implemented using measurement points at reclosers R1 and R4. If full selectivity is desired, additional measurement points can be deployed at intermediate reclosers (e.g., R1 to R4). In practice, a higher density of measurement points may be installed in areas with elevated wildfire risks to improve detection granularity and response speed.

For most practical scenarios, an open phase event produces a positive-sequence current variation exceeding a predefined threshold, allowing the use of positive-sequence connection impedance for detection. However, in cases where downstream $Y_g/\Delta$-connected transformers are unloaded and no DER is present, voltage restoration may occur and the resulting positive-sequence current variation may fall below the detection threshold. In such cases, if a detectable phase-current variation exists, phase-segregated connection impedance can be used as an alternative detection metric.

In closed-loop configurations with multiple utility sources, the calculated connection impedance is primarily determined by the overall line impedance rather than the equivalent load impedance. Consequently, for such applications, the monitored line sections should be selected to be sufficiently short, so that the connection impedance under an internal open phase condition remains significantly larger than the impedance of the monitored section. This ensures adequate sensitivity and reliable detection.

## VII. Conclusion

This paper presents a novel open phase detection method based on synchronized phasor measurements, in which the connection impedance between two measurement points is calculated for fault identification. Regardless of feeder topology, the calculated positive-sequence connection impedances under normal conditions, external or internal system events are closely approximated to the line impedance. Under an internal open phase condition in radial or open-loop distribution feeders, the calculated positive-sequence connection impedance is primarily approximated to the equivalent load impedance and is significantly larger than the line impedance. In addition, the calculated impedance exhibits a negative sign, which provides a reliable and distinctive indicator for open phase detection. For closed-loop distribution feeders with power sources from two directions, the calculated connection impedance is primarily influenced by the source impedance and the total line impedance. In such cases, a more sensitive impedance threshold is required to maintain detection effectiveness. DERs connected at the load side typically exhibit relatively high positive-sequence impedance and therefore do not adversely affect the proposed detection method. Although back-feed through downstream $Y_g/\Delta$-connected transformers can restore the open phase voltage, the proposed method remains effective provided that the positive-sequence current variation exceeds a minimum threshold, enabling reliable impedance calculation and fault detection.

The proposed open phase detection method requires only point-to-point communication (e.g., 4G or fiber-optic links) between two measurement points and does not necessitate significant additional infrastructure. It can be readily implemented using existing recloser controllers or RTUs. Furthermore, the method is inherently scalable and can operate with as few as two measurement points. Finally, the method provides inherent selectivity by detecting only internal open phase events within the monitored section, thereby minimizing the portion of the feeder that needs to be de-energized. This feature enhances both system reliability and operational safety.